\documentclass[twocolumn,prl,aps,showpacs]{revtex4}

\usepackage{psfrag,graphicx}

\usepackage{dcolumn}
\usepackage{amsmath,amssymb}
\usepackage{bm}
\usepackage[dvips]{color}

\definecolor{mygrey}{gray}{0.35}
\definecolor{mygreen}{rgb}{0.85,1,0.9}
\definecolor{myzard}{cmyk}{0,0,0.05,0}
\definecolor{mywhite}{rgb}{1,1,1}
\definecolor{myred}{rgb}{1,0,0}

\def\half{\textstyle\frac{1}{2}}

\def\fourth{\textstyle\frac{1}{4}}

\def\vec#1{{\bf{#1}}} 

 \def\ket#1{|#1\rangle}

\begin{document}

\title[Short Title]{
Quantum Chinos Game: 
winning strategies through quantum fluctuations}

\author{F. Guinea$^{\dag}$ and M.A. Mart\'{\i}n-Delgado$^{\ddag}$ 
} 
\affiliation{
$^{\dag}$Instituto de Ciencia de Materiales de Madrid,
CSIC, Cantoblanco, E-28049 Madrid, Spain.
\\ 
$^{\ddag}$Departamento de F\'{\i}sica Te\'orica I, Universidad Complutense,
28040. Madrid, Spain.  
}

\begin{abstract}
We apply several quantization schemes to simple versions of the Chinos game.
Classically,  for two players with one coin each, there is a symmetric
stable strategy that allows each player to win half of the times on average.
A partial quantization of the game (semiclassical) allows us to find a 
winning strategy for the second player, but it is unstable w.r.t. the
classical strategy. However, in a fully quantum version of the game
we find a winning strategy for the first player that is optimal:
the symmetric classical situation is broken at the quantum level.
\end{abstract}

\maketitle

In a typical scene at a Spanish restaurant, a small group of
companions-at-table gather at the bar extending their arms, each with their
clenched hands holding a few coins hidden inside. They are gambling for the
after-lunch round of coffees. One after another they tell a number, 
then open their hands showing their coins one another 
and count them all. Ofently, one of the pals smile meaning that s/he guessed 
the correct total number of coins. After a given number of plays, the player 
scoring the worst pays for all coffees. This  gambling game is known as
the {\em Chinos game} and  has been a traditional way in Spain
to decide who is in charge for the coffees' check\cite{chinos}. 

Interestingly enough, this simple-minded guessing game exhibits a rich 
variety  of patterns with complex behaviour that
has been used to model strategic behaviour in some social 
and economic problems, like financial markets and information 
transmission \cite{strategic}. This is an example of non-cooperative
game for each player seeks to maximize her/his chances of guessing
correctly, and at the same time to minimize the possibilities
of her/his opponents.

Recently, a new field for game theory has emerged in the form of 
quantum games with the goal of taking advantage of quantum effects
to attain a winning edge \cite{meyer}, \cite{ewm}, \cite{nature}. 
The blending of quantum mechanics with game theory opens novel strategies
based in exploiting the peculiarities of quantum behaviour, and
it has already estimulated a number of new ideas, e.g., in the
Prisioners' Dilemma there exists a quantum strategy that allows
both players to escape the dilemma\cite{ewm}.

In this letter our aim is twofold: firstly, to define quantum
versions of the Chinos game such that they reduce to the 
classical game as a limiting case. Secondly, to analyse the new
quantum versions in order to find how the classical strategies
behave under quantum effects, and if there exists new quantum
winning strategies without classical analogue.

\noindent{\em Classical Chinos Game.} In the classical formulation,
a number $N_{\rm p}$ of players enter the game, each having access
to $N_{\rm c}$ coins that they draw and hide in their hands at each
round of the game. Next, each player makes a guess about the total
number of coins held at that round,
with the constraint that no player can repeat the  number guessed
by the previous players.
Thus, the outcome of a given round may be either that one player wins
or failure for everyone.
As a remark, the heads and tails of the coins play no role in the 
Chinos game, so that they can be simply regarded as pebbles: 
only their number count. 

Let ${\cal D}:=\{0,1,\ldots,N_{\rm c}\}$ be the space of draws
and ${\cal G}:=\{0,1,\ldots,N_{\rm p}N_{\rm c}\}$ be the space
of guesses for the first player. 
Each players' movement has two parts: 1/ drawing coins; 2/
guessing the total number of coins altogether. 
Let us denote
by $M:=(d,g)$ one of these movements, with $d\in{\cal D}$ 
and $g\in\cal G$. The space of movements is 
${\cal M}:={\cal D}\times {\cal G}$ for the first player. 
Next players have a reduced 
guess space ${\cal G}'_{(i)}:={\cal G}-\{d_{(1)},...,d_{(i-1)}\},
i=2,...,N_{\rm p}$.
A possible strategy $S$
is an ordered sequence of movements $S:=(M_1,M_2,...,M_r)$
selected with some criteria or randomly, and
played during the $r$ rounds that the whole game takes.

We shall denote by 
CCG($N_{\rm p},N_{\rm c}$) a classical Chinos game of 
$N_{\rm p}$ players and $N_{\rm c}$ coins. The exhaustive analysis
of such a generic game turns out to be too complicated 
\cite{strategic}, thus we shall concentrate on the case of only
$N_{\rm p}=2$ players for which we have the following result:

{\em $1^{st}$ Result}. Let us denote the classical strategies
for each player $i=1,2$ by 
$S_{(i)}:=(M_{(i),1},M_{(i),2},...,M_{(i),r})$. Then, the best strategy for
player 1 is to choose movements $M_{(1),j}, j:=1,2,...,r$ with
$d_{(1),j}$ randomly distributed and $g_{(1),j}=N_{\rm c}, \forall j$,
while the best strategy for player 2 is to choose draws $d_{(2),j}$
at random. For $r$ large enough, the result of the game is even.
 
{\em Proof.} Since the Chinos game is a non-cooperative game,
in this result we are assuming that one of the main goals of player 1
is not to transmit any information to player 2 about her/his values
$d_{(1),j}$. This can be achieved by choosing $g_{(1),j}=N_{\rm c}$
irrespective of the number that s/he draws. Moreover, players soon realize
that  as they cannot know in advance her/his opponent strategy, the best
strategy they can choose is to pick $d_{(i),j}, i=1,2;\forall j$ at random.
Now, let us call $p_1$ the probability that player 1 guesses correctly
the total sum they are after, namely, $a_j:=d_{(1),j}+d_{(2),j}$, and
similarly for $p_2$. The quantities each player is interested in maximizing
are the normalized probabilities $P_i:=p_i/\sum_{i=1,2}p_i$. Thus,
under these circumstances, the probability that the second player guesses
the correct sum is 
\begin{equation}
p_2 = \frac{1-p_1}{N_{\rm c}}.
\label{qc1}
\end{equation}
Then, the quantity player 2 wants to optimize is
\begin{equation}
P_2 = \frac{1-p_1}{1+p_1(N_{\rm c}-1)},
\label{qc2}
\end{equation}
which is a decreasing function of $p_1$, so that player 2 is interested in
reducing $p_1$ as much as possible. However, player 1 can always resort to
make random guesses about the number of coins drawn by player 2. This amounts
to a lowest bound on $p_1$ given by $p_{1,<}:=1/(N_{\rm c}+1)$. Therefore,
player 2 should draw coins at random so that $p_1$ cannot exceed $p_{1,<}$
and we end up with an even situation given by
\cite{ccg2}
\begin{equation}
P_1 = P_2 = \frac{1}{2}.
\label{qc3}
\end{equation}
$\blacksquare$

We may view this result as a sort of ``classical symmetry''
between players 1 and 2:
\begin{equation}
{\rm Player} \ 1 \longleftrightarrow {\rm Player} \ 2,
\label{qc3b}
\end{equation}
in the sense that there is now way to untight the result of
the game if both players play at random. Our goal is to 
define quantum extensions of the Chinos game to see if this
symmetry can be broken at the quantum level.
We shall use this classical result as a guide to analyse the 
behaviour of classical strategies when we enter the realm of the
quantum extensions of the game.

\noindent{\em Semiclassical Chinos Game.} A first attempt at quantizing the
Chinos game is to make a quantum extension of the space of draws ${\cal D}_q$
while leaving the space of guesses ${\cal G}$ classical. We term this case
semiclassical for obvious reasons and denote by SCG($N_{\rm p},N_{\rm c}$) a
semiclassical Chinos game. The natural choice for ${\cal D}_q$ is
to replace coins by quantum coins or qubits. Likewise, a quantum two-level
system is represented by a spin $\half$ particle with states 
$\ket{\!\uparrow},\ket{\!\downarrow}$ representing heads and tails, 
respectively. However, we find that spins are not appropriate in the Chinos
game since only the presence or absence of coins in players' hands matters.
Hence, a more suitable way of representing qubits is to use a
boson system defined by bosonic creation/annihilation operators 
$b^{\dag},b$ obeying canonical commutation relations (CCR) $[b,b^{\dag}]=1$
and acting on the bosonic vacuum $\ket{0}$ in the standard fashion:
$b\ket{0}=0, b^{\dag}\ket{n}=\sqrt{n+1}\ket{n+1}$, with 
$\ket{n}:=(b^{\dag})^n\ket{0}/\sqrt{n!}$.

For simplicity, we shall consider first the case in which each quantum
player has only one coin, namely, SCG($N_{\rm p},1$).

To each player $i=1,2,...,N_{\rm p}$ we shall assign a set of operators
$O(\theta_i,\phi_i)$ parameterized by the two angles characterizing a qubit
state in the Bloch sphere. Thus, we introduce
\begin{equation}
O_i(\theta,\phi):=\cos\half\theta_i+
{\rm e}^{{\rm i}\phi_i}\sin\half\theta_i\ b^{\dag},
\ \ 2\theta_i,\phi_i\in[0,2\pi).
\label{qc4}
\end{equation}
These operators represent the quantum draw space ${\cal D}_q$.
At a given round $j$ of the game, each player selects one possible operator
$O_i(\theta,\phi)$ and at the end of the drawing process, we represent the
situation of having all players' hands together
by the following joint quantum state
\begin{equation}
\ket{\Psi_{\rm SCG}^{(N_{\rm p},1)}}:= N^{-1/2}
\prod_{i=1}^{N_{\rm p}}O_i(\theta,\phi)\ket{0}
=\sum_{n=0}^{N_{\rm p}}c_n\ket{n},
\label{qc5}
\end{equation}
where $N$ is a normalization constant and $c_n$ expansion coefficients.
This state faithfully represents the fact that what really counts
is to guess the total sum $a_j=\sum_{i=1}^{N_{\rm p}}d_{(i),j}\in {\cal G}$,
no matter what the partial contributions $d_{(i),j}$ of each player are.
Moreover, the quantum effects are clearly apparent since when the state 
$\ket{\Psi_{\rm SCG}^{(N_{\rm p},1)}}$ is expanded in states 
$\ket{n}, n\in {\cal G}$, each coefficient $c_n$ receives contributions
from each player that cannot be factorized out. Then, with (\ref{qc5})
we can compute the probability $p(n)$ that any player obtains the value
$g=n$ after a measurement, namely,
\begin{equation}
p(n):=|\langle n\ket{\Psi_{\rm SCG}^{(N_{\rm p},1)}}|^2=c_n^2 n!
\label{qc6}
\end{equation}

With the present quantization scheme we have an infinitely many number of
possible draws. In practice, it is a reasonable assumption to reduce 
the possible operator choices to a finite restricted set. To be concrete,
let us consider the case of $N_{\rm p}=2$ players SCG${(2,1)}$ 
and we select from (\ref{qc5}) the following reduced operator set 
\begin{equation}
\begin{split}
& O_1:= I, O_2:= \frac{1}{\sqrt{2}} (I+b^{\dag}), \\
& O_3:= \frac{1}{\sqrt{2}} (I-b^{\dag}), O_4:=b^{\dag}.
\end{split}
\label{qc7}
\end{equation}
Notice that operators $O_1$ and $O_4$ represent the classical draws of
0 and 1, respectively, while $O_2$ and $O_3$ represent novel quantum
superpositions of the classical draws.
These conditions represent a generic situation to analyse quantum effects
in the Chinos game and we find the following result:

{\em $2^{nd}$ Result}. i) The strategy of drawing randomly 
from (\ref{qc7})
becomes a winning strategy for player 2. However, this strategy
is unstable. ii) The classical strategy of drawing randomly between
$O_1$ and $O_4$ is a winning strategy for both players (evenness)
and is stable.

{\em Proof.} The analysis relies on Table~\ref{tableSCG1} showing the
probabilities of obtaining 0,1 and 2 coins when player 1 draws
operator $O_{i_1}^{(1)}$ and player 2  draws $O_{i_2}^{(2)}$, 
$i_1,i_2 = 1,2,3,4$, according
to (\ref{qc5})-(\ref{qc7}). i/ Let us assume that players 1 and 2 both
know the classical winning strategy of a CCG and decide to make a 
straightforward generalization of it to the semiclassical case SCG.
Then, player 2 decides to play random draws among the four possible choices
in (\ref{qc7}). In this situation, player 1 is left with a set of probabilities
of getting a number of coins 0,1 and 2 given by Table~\ref{tableSCG2},
which are computed from Table~\ref{tableSCG1}
by tracing out (averaging) over player 2. Hence, 
if the second player plays at random, the best choice for
player 1 is to guess 1 (or 0) if s/he draws  $O^{(1)}_1$, and 2
if s/he draws  $O^{(1)}_2$,$O^{(1)}_3$ and $O^{(1)}_4$. 
Thus, her/his total chances of winning are:
\begin{equation}
P_1 = \fourth  \times \frac{1}{2} + \half \times
\frac{68}{168} + \fourth \times \frac{7}{12} = \frac{53}{112} < \half
\label{qc8}
\end{equation}
Therefore, the strategy of both players drawing at random is no
longer an even strategy in this case.

\begin{center}
\begin{table}[tH]
\begin{tabular}{||c||c|c|c|c||}
\hline \hline
&$O^{(1)}_1$ &$O^{(1)}_2$ &$O^{(1)}_3$ &$O^{(1)}_4$ \\
\hline \hline
$O^{(2)}_1$
&$\begin{array}{c}p(0) = 1 \\ p(1) = 0 \\ p(2) = 0 \end{array}$ 
&$\begin{array}{c}p(0) = \frac{1}{2} \\ p(1) = \frac{1}{2}
\\ p(2) = 0 \end{array}$ 
&$\begin{array}{c}p(0) = \frac{1}{2} \\ p(1) = \frac{1}{2}
\\ p(2) = 0 \end{array}$ 
&$\begin{array}{c}p(0) = 0 \\ p(1) = 1
\\ p(2) = 0 \end{array}$  \\ \hline
$O^{(2)}_2$
&$\begin{array}{c}p(0) = \frac{1}{2} \\ p(1) = \frac{1}{2}
\\ p(2) = 0 \end{array}$ 
&$\begin{array}{c}p(0) = \frac{1}{7} \\ p(1) = \frac{4}{7}
\\ p(2) = \frac{2}{7} \end{array}$ 
&$\begin{array}{c}p(0) = \frac{1}{3} \\ p(1) = 0
\\ p(2) = \frac{2}{3} \end{array}$ 
&$\begin{array}{c}p(0) = 0 \\ p(1) = \frac{1}{3}
\\ p(2) = \frac{2}{3} \end{array}$  \\ \hline 
$O^{(2)}_3$
&$\begin{array}{c}p(0) = \frac{1}{2} \\ p(1) = \frac{1}{2}
\\ p(2) = 0 \end{array}$ 
&$\begin{array}{c}p(0) = \frac{1}{3} \\ p(1) = 0
\\ p(2) = \frac{2}{3} \end{array}$ 
&$\begin{array}{c}p(0) = \frac{1}{7} \\ p(1) = \frac{4}{7}
\\ p(2) = \frac{2}{7} \end{array}$ 
&$\begin{array}{c}p(0) = 0 \\ p(1) = \frac{1}{3}
\\ p(2) = \frac{2}{3} \end{array}$ \\ \hline 
$O^{(2)}_4$
&$\begin{array}{c}p(0) = 0 \\ p(1) = 1
\\ p(2) = 0 \end{array}$
&$\begin{array}{c}p(0) = 0 \\ p(1) = \frac{1}{3}
\\ p(2) = \frac{2}{3} \end{array}$
&$\begin{array}{c}p(0) = 0 \\ p(1) = \frac{1}{3}
\\ p(2) = \frac{2}{3} \end{array}$
&$\begin{array}{c}p(0) = 0 \\ p(1) = 0
\\ p(2) = 1 \end{array}$ \\ \hline \hline 
\end{tabular}
\caption{Probabilities for the outcomes of total coins 0,1 and 2 in
a SCG($2,1$) game. In the horizontal, the draws of player 1 and in the
vertical, the draws for player 2.}
\label{tableSCG1}
\end{table}
\end{center}
ii/ However, after a large number of rounds $r$, player 1 will realize
that playing at random is a winning strategy for her/his opponent and
then s/he will seek to improve it. To do this, s/he may resort to draw
only the classical choices . Then, from Table~\ref{tableSCG2}, her/his
chances of winning are
\begin{equation}
P_1 = \half  \times \half + \half \times
\frac{7}{12} = \frac{13}{24} > \half
\label{qc9}
\end{equation}
This implies that the strategy in i) is not stable.
Likewise, player 2 will not be happy with this new situation.
S/he will try to match player's 1 strategy by choosing the 
same purely classical strategy. This fully classical situation
is represented by the boxes at the outer corners of Table~\ref{tableSCG1}.
Then we are led to $P_1=P_2=\half$ as the stable best strategy
as in (\ref{qc3}).

\noindent $\blacksquare$

This result means that if player 1 applys her/his knowledge of the 
classical game naively by drawing at random from the four choices available,
in the long run s/he will realize that player 2 gets a winning edge.

\noindent{\em Quantum Chinos Game.} Motivated by the previous semiclassical
analysis, we propose a fully quantized version of the Chinos game by
quantizing both the draw space ${\cal D}_q$ 
and the guessing space ${\cal G}_q$. 
We shall define the quantum space of guesses ${\cal G}_q$ by allowing
each player to make a guess about the final quantum state 
$\ket{\Psi_{\rm SCG}^{(N_{\rm p},1)}}$
in (\ref{qc5}), and not merely
about the possible outcomes of the total coins. Thus, 
each player $i$ will make a guess 
$\ket{\Psi_i}, i=1,2,...,N_{\rm p}$ about what the actual joint quantum
state they are dealing with. Moreover, we also extend quantumly
the classical constraint that the guess $g_i$ of player $i$ cannot
be the same as guesses $g_j$ for $i<j$. This is achieved by imposing
that the guess a player $i$ can make is restricted to the subspace
orthogonal to the space spanned by the guesses of the previous players,
namely,
\begin{equation}
{\cal G}_{q,i}:={\rm span}\{\ket{\Psi_1},...,\ket{\Psi_{i-1}}\}^{\perp}.
\label{qc10}
\end{equation}

\begin{center}
\begin{table}
\begin{tabular}{||c|c|c|c||}
\hline \hline $O^{(1)}_1$ &$O^{(1)}_2$ &$O^{(1)}_3$ &$O^{(1)}_4$ \\
\hline \hline $\begin{array}{c}\langle p(0)\rangle = \frac{1}{2} \\
\langle p(1)\rangle = \frac{1}{2} \\ \langle p(2)\rangle = 0 \end{array}$
&$\begin{array}{c}\langle p(0)\rangle = \frac{41}{168} \\ \langle
p(1)\rangle = \frac{59}{168} \\ \langle p(2)\rangle = \frac{68}{168} \end{array}$
&$\begin{array}{c}\langle p(0)\rangle = \frac{41}{168} \\ \langle
p(1)\rangle = \frac{59}{168} \\ \langle p(2)\rangle = \frac{68}{168} \end{array}$
&$\begin{array}{c}\langle p(0)\rangle = 0 \\ \langle p(1)\rangle =
\frac{5}{12} \\ \langle p(2)\rangle = \frac{7}{12} \end{array}$ \\ \hline \hline
\end{tabular}
\caption{Averaged probabilities of obtaining 0,1, and 2 coins by 
player 1 in a SCG($2,1$) game according to the draws $O_i, i=1,2,3,4$ s/he makes. }
\label{tableSCG2}
\end{table}
\end{center}
With these new rules, we need to define a new function payoff:
the gain for player $i$ is
\begin{equation}
f_i:=|\langle \Psi_i\ket{\Psi_{\rm SCG}^{(N_{\rm p},1)}}|^2.
\label{qc11}
\end{equation}
This way of quantizing the space of guesses is reminiscent of the theory
of quantum algorithms \cite{rmp}, and more specifically, from quantum
searching algorithms \cite{grover1}, \cite{meyer2}.
That this fully quantum version of the Chinos game includes the classical
one is guaranteed since the latter appears as a particular case when the
only allowed guesses are $\ket{0},\ket{1},...,\ket{N_{\rm p}}$ (if the
number of coins per player is $N_{\rm c}=1$.)

For simplicity, we shall consider the quantum case QCG($N_{\rm p},N_{\rm c}$)
for two players and one coin each, and their quantum guesses 
comprise the finite
set (\ref{qc7}). We find the following result:

{\em $3^{rd}$ Result}. In a quantum Chinos game CCG($2,1$), the first 
player has a stable winning strategy that allows her/his to win more
than half of the times.

{\em Proof.} A systematic analysis in this case of 2 players with
one coin each proceeds as follows. The space of draws for player 1
is ${\cal D}_{q}$ given by the reduced set (\ref{qc7}). Then,
player 1 draws $O^{(1)}_{i_1}\in {\cal D}_{q}$. The space of guesses
for player 1 is ${\cal G}_{q,1}:=\{ O^{(1)}_{j_1} O^{(1)}_{k_1}, 
1\leq j_1 \leq k_1 \leq 4\}$. Next, player 1 makes a quantum guess
$g_{q,1}:=O^{(1)}_{j_1} O^{(1)}_{k_1}\in {\cal G}_{q,1}$. 
Now enters player 2 with a draw $O^{(2)}_{i_2}\in {\cal D}_{q}$,
and making a guess $g_{q,2}:=O^{(2)}_{j_2} O^{(2)}_{k_2}$ that in order
to be eligligible, it has to be orthogonal to player's 1 guess
(\ref{qc10}). To characterize this orthogonality condition, it is convenient
to introduce the following $16\times 16$ matrix 
\begin{equation}
\begin{split}
G_{(j_1,k_1),(j_2,k_2)}:= &
\frac{\langle0|O^{\dag}_{j_1} O^{\dag}_{k_1}
O^{}_{j_2} O^{}_{k_2}\ket{0}}{\sqrt{N_{j_1k_1}^{}}\sqrt{N_{j_2k_2}^{}}}
,\\
N_{jk}^{}:=& \langle0|O^{\dag}_{j} O^{}_{k}\ket{0},
\label{qc11b}
\end{split}
\end{equation}
which can be thought of as a metric on the quantum guess space.
Thus, guess $g_{q,2}$ is admissible for the given guess $g_{q,1}$
iff $G_{(j_1,k_1),(j_2,k_2)}=0$. Finally, for a pair of draws,
the actual joint state representing that round of the game is
\begin{equation}
\ket{\Psi_{\rm QCG}^{(2,1)}} = 
N_{12}^{-1/2}O^{(1)}_{i_1} O^{(2)}_{i_2}\ket{0},
\label{qc11c}
\end{equation}
while the function payoffs $f_i,i=1,2$ for each player can also be read off
from the metric (\ref{qc11b}) as follows
\begin{equation}
\begin{split}
f_1 = |G_{(j_1,k_1),(i_1,i_2)}|^2,\\
f_2 = |G_{(j_2,k_2),(i_1,i_2)}|^2.
\label{qc11d}
\end{split}
\end{equation}
Then, once we have computed
the metric (\ref{qc11b}), it is possible to make an exhaustive
study of all the possibilities in this quantum Chinos gain and compute
each players' payoffs for each of those possibilities. We have
performed this analysis with the following result:
let us show that if player 1 makes  draws 
with equal probability among the
choices $O^{(1)}_2$ and  $O^{(1)}_3$ only (\ref{qc7}), then
s/he is half-way for a winning position. The rest of the strategy
is to set up the quantum guesses as follows. 
When player 1 draws $O^{(1)}_2$, s/he decides to 
make always the following quantum guess
\begin{equation}
\ket{\Psi_1}:=N_1^{-1/2} O^{(1)}_2 O^{(1)}_2 \ket{0}=
\frac{1}{\sqrt{7}}(\ket{0}+2\ket{1}+\sqrt{2}\ket{2}),
\label{qc12}
\end{equation}
in which case, a possible guess for player 2 
according to (\ref{qc10}) would be
\begin{equation}
\ket{\Psi_2}:=N_2^{-1/2} O^{(2)}_3 O^{(2)}_4 \ket{0}=
\frac{1}{\sqrt{3}}(\ket{1}-\sqrt{2}\ket{2}).
\label{qc12b}
\end{equation}
While if s/he draws $O^{(1)}_3$, s/he decides to 
make always the following quantum guess
\begin{equation}
\ket{\Psi_1}:=N_1^{-1/2} O^{(1)}_3 O^{(1)}_3 \ket{0}=
\frac{1}{\sqrt{7}}(\ket{0}-2\ket{1}+\sqrt{2}\ket{2}).
\label{qc13}
\end{equation}

Now, let us analyse the case when player 1 draws $O^{(1)}_2$.
Then, player 2 is left with the four draws in the set (\ref{qc7})
and the correspoding joint final states $\ket{\Psi_{\rm SCG}^{(2,1)}}$
that we collet in Table~\ref{tableSCG3}. When the first player
draws $O^{(1)}_3$, then we obtain a similar table by exchanging
$2 \leftrightarrow 3$.

>From Table~\ref{tableSCG3} we see that under these circumstances,
it is clear that player 2 will avoid to make the classical draws
$O^{(2)}_1$ and $O^{(2)}_4$, since they yield payoffs
$f_1=\frac{9}{14}>\half$, $f_1=\frac{16}{21}>\half$
for the first player. Thus, player 2 is led to play only the draws
$O^{(2)}_2$ and $O^{(2)}_3$ at random. However, even in this case,
player 1 will have a winning edge on the average since the chances
of winning for the first player are
\begin{equation}
\langle f_1\rangle = \half \times 1 + \half \times \frac{1}{21} = \frac{11}{21} > \half.
\label{qc14}
\end{equation}
\begin{table}[th]
\begin{tabular}{||c|c|c||}
\hline\hline
Quantum guess &Joint state $\ket{\Psi_{\rm CCG}^{(2,1)}}$&Gain for player 1  \\ \hline\hline
$O^{(2)}_1$ & $\frac{1}{\sqrt{2}} \left( | 0 \rangle +
| 1 \rangle \right)$ &$f_1=\frac{9}{14}$ \\ \hline
$O^{(2)}_2$ & $\frac{1}{\sqrt{7}} \left( | 0 \rangle +
2 | 1 \rangle + \sqrt{2} | 2 \rangle \right)$ & $f_1=1$ \\ \hline
$O^{(2)}_3$ & $\frac{1}{\sqrt{3}} \left( | 0 \rangle -
\sqrt{2} | 2 \rangle \right)$ &$f_1=\frac{1}{21}$ \\ \hline
$O^{(2)}_4$ &  $\frac{1}{\sqrt{3}} \left( | 1 \rangle +
\sqrt{2} | 2 \rangle \right)$ &$f_1=\frac{16}{21}$ \\ \hline\hline
\end{tabular}
\caption{ Quantum guesses for
player 2 when player 1 draws $O^{(1)}_2$ (\ref{qc7}), and the corresponding
joint state (\ref{qc11c}) 
and gains for player 1 (\ref{qc11d}).}
\label{tableSCG3}
\end{table}
\noindent $\blacksquare$

{\it Conclusions.}
In game theory, players strive for even the slightest advantage
that would tilt a game's outcome in their favor.
We have found that the chances of winning for player 1 are better off
on average than those of her/his opponent. We may interpret this result
as the breaking of the symmetric classical situation (\ref{qc3b}) at the
quantum level:
\begin{equation}
{\rm Player}\ 1 \longleftrightarrow \hskip -14pt / \hskip 9pt {\rm Player} \ 2.
\label{qc15}
\end{equation}
This advantage of the first player resembles a similar situation found
in the PQ quantum game\cite{meyer}. In the present case, however,
the correlation between players in the final result is dynamically generated, 
i.e., it is a consequence of the player's choice, and it is not encoded
in the initial state. In this respect, it also differs from
the quantum generalization of other simple games, like
the prisoner's dilemma\cite{ewm}, or the minority game\cite{KJB01}.

\noindent {\em Acknowledgments}. We acknowledge financial support
from projects: PGC96-0875, 07N/0015/2001 and PB98-0685.


\begin{references}
\bibitem{chinos}
We do not claim that the Chinos game originates from Spain, but
the word chinos is a deformation of chinas meaning pebbles in
Spanish.
\bibitem{ccg2}
If player 1 gives away some information about the
coins s/he draws to player 2, then 
$p_2>\frac{1-p_1}{N_{\rm c}}$ and 
$P_1<\frac{N_{\rm c}p_1}{1-p_1(N_{\rm c}-1)}$.
As player 1 wants $P_1>\half$, then it follows that
$p_1>\frac{1}{N_{\rm c}+1}$. On the contrary, player
2 wants to minimize $P_1$ by drawing coins at random
so that $p_1=\frac{1}{N_{\rm c}+1}$, bringing the 
situation back to the case (\ref{qc3}).
\end{references}
\end{document}